\documentstyle[prb,aps,preprint]{revtex}
\begin{document}
\draft
\title{Zone edge focused two-phonon processes in He atom scattering from a simple prototype system: Xe(111).}
\author{Antonio \v{S}iber and Branko Gumhalter\footnote{Senior Research Associate of the {\em Abdus Salam} International Centre for Theoretical Physics, Trieste, Italy}\thanks{Corresponding author. E-mail: branko@ifs.hr , tel. +385-1-4698805, fax +385-1-4698890}}
\address{Institute of Physics of the University, P.O.\ Box 304,
10001 Zagreb, Croatia. }
\maketitle

\def\bcalR{{\hbox{\boldmath$\cal{R}$}}}

\begin{abstract}
We present a theoretical analysis of the multiphonon features appearing in the angular and energy resolved scattering spectra of low energy He atoms inelastically scattered from surface phonons. By applying the recently developed fully quantum multiphonon scattering formalism to the benchmark system Xe(111) 
we are able to reproduce with great accuracy the experimental data and also demonstrate how in the transition from a single- to a few- to a multiphonon scattering regime the one- and the many-phonon components of the scattering spectra evolve as the mean number of exchanged phonons increases with the increase of He atom incident energy.  
We show how the two-phonon processes may give rise to distinct peaks in the scattering spectrum as a result of the zone edge focusing effects and how the many-phonon features may still produce a structured background before the true multiphonon limit characterised by a Gaussian spectral shape is reached. These findings should prove useful in the interpretation of multiphonon He atom scattering spectra from other systems with similar surface vibrational properties. 

\end{abstract}

\newpage


The energy and momentum transfer in gas-surface collisions proceeds under standard conditions dominantly in the multiphonon scattering regime. A convenient laboratory tool for studying these processes is inelastic scattering of thermal energy He beams from surfaces either in the ultra-high vacuum (UHV) environment, which enables to achieve and control the cleanliness of the studied surface on the atomic scale by means of the various surface science techniques, or in the wind 
tunnel experiments in which the energy transfer is directly obtainable from calorimetric measurements performed on samples exposed to the beam\cite{Legge}. However, the structure and cleanliness of the sample surface in the wind tunnel environment can be controlled and maintained only at a technical level.  

The majority of UHV experiments utilising low energy He-atom-scattering time-of-flight (HAS-TOF) spectroscopy of surfaces have been performed with the aim to investigate and reveal the properties of surface phonons. To this end the experiments are carried out in the one-phonon scattering regime as only such experimental conditions allow clear fingerprinting of the surface phonon dispersion curves which serve as prerequisites for mode assignments and theoretical interpretations of the spectral peak intensities\cite{Hulpke}.
The interpretation of HAS experiments in the multiphonon scattering regime is much more involved because of the need to deconvolute the various features in HAS-TOF spectra whose energies and intensities depend on phonon dispersions and scattering dynamics in a highly nontrivial fashion. Due to this complication very few experiments carried out in the multiphonon scattering regime have been complemented by adequate theoretical analyses, except in the case of high enough He beam energies for which the HAS-TOF spectra acquire a structureless Gaussian shape typical of the semiclassical scattering limit\cite{comment,Hofmann}. A particularly complicated situation regarding the interpretation of measurements arises in the transition or cross-over from a single to a few-phonon scattering regime because of the difficulties encountered in theoretical assessments of the scattering events of this kind.

An exception to the general situation characterised by the difficulties to study transitions from a single- to a few- to a multiphonon scattering regime occurs for surfaces which exhibit nondispersive or Einstein-like phonons whose frequency $\omega({\bf Q})$ as a function of a two-dimensional phonon wavevector ${\bf Q}$ parallel to the surface remains constant over the surface Brillouin zone (SBZ), i.e. $\omega({\bf Q})=\omega_{0}$. Typical examples of the systems exhibiting nondispersive surface phonons are monolayers of rare gas atoms adsorbed on low index crystal surfaces of metals\cite{sibener,Xemulti} and layers of molecular adsorbates like CO on metals\cite{WitteCO,Witte}.
The HAS TOF spectra originating from single and multiphonon excitation of Einstein modes display a relatively simple structure consisting of a series of equidistant peaks located at multiples of Einstein phonon energy $\hbar\omega_{0}$. 
The intensities of higher order peaks increase as the scattering regime changes from a single- to a multiphonon one but the actual transition between the two regimes depends on several parameters such as the incident energy and angle of the projectile atom, the magnitude of the mode frequency and the direction of its polarisation, surface temperature etc. 

The quantity that best reflects the transitions between different scattering regimes is the exponent $2W$ of the Debye-Waller factor (DWF), $\exp(-2W)$, that is calculated for a particular scattering spectrum. The value of $2W=\bar{n}$ is a function of the afore mentioned parameters and measures the mean number of phonons $\bar{n}$ exchanged in the course of collision\cite{SRL}. Hence, the transition from a single to a multiphonon scattering regime occurs for those values of the system and collision parameters for which $2W\simeq 1$. In the case of nondispersive Einstein phonons this value of the Debye-Waller exponent (DWE) gives rise to maximum intensity of the first Einstein phonon peak in the scattering spectrum, whereas for $2W\geq 2$ the intensity of higher order peaks takes over. 
The recently developed quantum theory of multiphonon HAS from nondispersive phonons\cite{recovery,plrep}, which enables a detailed analysis of the properties of the DWF\cite{DWF}, has been successfully applied to interpret the TOF spectra of He atoms scattered from monolayers of Xe atoms on Cu substrates\cite{Xemulti} in which the dominant inelastic structure arises from excitation of Einstein modes. Many conclusions reached in these analyses can be conveniently extended also to the analyses of more complicated situations of inelastic HAS from dispersive phonons.    

The problem of observation and pinpointing of few- and many-phonon processes in inelastic HAS from dispersive phonons is much more critical than in the case of Einstein modes and until recently no clearcut experimental evidence as to their resolution in the HAS TOF spectra was available.
However, recent high resolution HAS studies of phonons characteristic of the (111) surface of fcc van der Waals crystal of xenon atoms condensed on Pt(111) substrate\cite{Xe111} have revealed some distinct features in the TOF spectra which could not be attributed to the single phonon peaks over a structureless background. 
The dynamical matrix analysis of the Xe(111) surface\cite{Xe111} enabled a clear assignment of the Rayleigh wave (RW), the longitudinal resonance (LR) and the shear horizontal (SH) phonons in the TOF spectra but not of the extra features which for incident He atom energies of $E_{i}=10.5$ meV and small parallel momentum transfer $\Delta{\bf K}$ showed up as discernible maxima at energy loss $\Delta E\simeq -4.2$ meV. 

To interpret the experimental results we have carried out  fully quantum multiphonon calculations of the energy and parallel momentum resolved scattering spectra 
$N_{{\bf k_{i}},T_{s}}(\Delta E,\Delta{\bf K})$ for experimentally fixed 
He atom initial momentum $\hbar{\bf k_{i}}$ and the substrate temperature $T_{s}$, 
using the exponentiated Born approximation (EBA) formalism\cite{plrep,HAS}. In this formalism the scattering spectrum as the function of the energy and parallel momentum transfer from the projectile to the phonons is given by a compact expression:

\begin{equation}
N^{EBA}_{{\bf k_{i}},T_{s}}(\Delta E,\Delta{\bf K}) =
e^{-2W}
\int^{\infty}_{-\infty} \frac{d\tau d^{2}{\bcalR}}{(2\pi\hbar)^{3}}
e^{\frac{i}{\hbar}[(\Delta E)\tau-\hbar(\Delta{\bf K}){\bcalR}]}
\exp[2W_{{\bf k_{i}},T_{s}}^{EBA}(\bcalR,\tau)], 
\label{eq:specEBA}
\end{equation}
where the Debye-Waller exponent is $2W=2W_{{\bf k_{i}},T_{s}}^{EBA}(0,0)$ and the scattering function $2W_{{\bf k_{i}},T_{s}}^{EBA}(\bcalR,\tau)$ is calculated in the EBA (for details see Eq. (12) of Ref. \onlinecite{Xe111}). The values of $\Delta E$ and $\Delta{\bf K}$ appearing in expression (\ref{eq:specEBA}) are not independent but are connected by the requirement of total energy and parallel momentum conservation in the collision, i.e. they are confined to the scan curve.  
 
The calculated values of the Debye-Waller exponents pertaining to the measured spectra reported in Ref. \onlinecite{Xe111} (c.f. Fig. 12 of Ref. \onlinecite{Xe111}) indicated that these experiments were 
carried out in the two-phonon scattering regime. A typical result of the EBA calculations for the scattering spectrum of the He$\rightarrow$Xe(111) collision system and comparison with the experimental data reported in Ref. \onlinecite{Xe111} is shown in Fig. \ref{vas10_1}. 
Under these scattering conditions the observed features at $\Delta E=\pm$ 4.2 meV, which can not be interpreted as arising from single phonon scattering processes, can be explained within the present model as a result of excitation of two RW phonons with wavevectors from nearly the opposite edges of the first SBZ so as that the total momentum transfer to phonons is small. A similar effect has also been observed in inelastic HAS from frustrated translation modes of c$(2\times 2)$ structure of CO on Cu(100)\cite{Witte}.
Following such a good agreement between the measured and theoretical results and taking into account the information on the scattering regime embodied in the corresponding DWE, we find a strong support for the above proposed assignments of the extra features in the He$\rightarrow$Xe(111) TOF spectra. The goal of the present work is to further corroborate this interpretation through supplementing the previous calculations by more detailed theoretical analyses of the few-phonon scattering processes which may occur in HAS studies of this and other systems exhibiting similar surface vibrational properties. 

In Fig. \ref{vas10_2} we display four scattering spectra calculated for the collision system He$\rightarrow$Xe(111) 
using the theoretical formalism outlined in Sec. V of Ref. \onlinecite{Xe111}. The parameter which is varied in calculations is the incident energy $E_{i}$ of the He atom whereas all other parameters are kept fixed (incident angle, surface temperature, etc). 
Since the maximum energies of the RW, LR and SH phonons that are localized at the Xe(111) surface are of the order of few meV only (see Fig. 11 of Ref. \onlinecite{Xe111}), the probability of multiple excitation of surface phonons is non-negligible already for the lowest studied energy $E_{i}=5.7$ meV for which the scattering spectrum in the upper left-hand-side (LHS) panel of Fig. \ref{vas10_2} has been calculated. 
This is  
corroborated by the magnitude of the corresponding Debye-Waller exponent $2W=1.7$ which implies that a part of the spectral weight has been shifted from the single to the multiphonon structure. The spectrum is dominated by the peaks signifying the excitation and annihilation of single RW and LR phonons in the scattering event but some structured multiphonon background is already discernible (dashed line) both under the one-phonon peaks and in the region between them. 
This multiphonon structure, and in the case studied this is a two-phonon structure, becomes more pronounced as the incoming energy is increased to $E_{i}=10.4$ meV for which $2W=2.2$, as shown in the upper right-hand-side (RHS) panel in Fig. \ref{vas10_2}. The peaks denoted by "M" are of purely multiphonon origin but depending on their position they reflect kinematically different processes. The one at zero energy transfer $\Delta E=0$ and denoted by M$_{0}$ arises as a consequence of creation and annihilation of several phonons with zero total energy but nonzero total momentum balance. 
Note that the processes of this kind are possible only for finite substrate temperature and their effect in producing the high intensity of the no-loss or elastic line is particularly noticeable in the case of nondispersive phonons, as has been discussed in Refs. \onlinecite{recovery} and \onlinecite{plrep}.

 The origin of the peaks M$_{2}$ and M$_{-2}$ located at $\Delta E\pm\simeq 4.2$ meV, respectively, is more subtle as it depends on a strong interplay between the characteristics of the phonon density of states typical of the studied system and the dynamics of the scattered particle. The RW phonons whose polarisation is dominantly perpendicular to the surface couple most strongly to the scattered He atoms, and this property holds in the single as well as in the multiphonon scattering regime. Since both the perpendicular polarization and the RW phonon density of states reach maximum at the SBZ boundary, it is expected that the largest contributions to the two-phonon processes will come from excitations of two Rayleigh waves with respective wavevectors ${\bf Q_{1}}$ and ${\bf Q_{2}}$ from the vicinity of the zone edges, i.e. $\mid{\bf Q_{1}}\mid =\mid{\bf Q_{2}}\mid\simeq Q_{max}$.
On the other hand, as the one-phonon scattering matrix elements also depend on the projectile dynamics and produce maximum two-phonon intensity for minimum total momentum transfer\cite{plrep,Xe111,HAS}, i.e.  $\mid\Delta {\bf K}\mid=\mid{\bf Q_{1}+Q_{2}}\mid= min$, the combination of two requirements
 leads to the condition that maximum two-phonon scattering intensity occurs if the phonon wavevectors are pointed to the nearly opposite edges of the SBZ.
 Such constrained or "zone edge focused" two phonon creation and annihilation processes give rise to the peaks in the scattering spectra that are located approximately at twice the value of the zone edge frequency of a single excited phonon. This is clearly visible in both upper panels of Fig. \ref{vas10_2} in which the peaks M$_{\pm 2}$ indeed appear at $\sim \pm 4.2$ meV, i.e. at twice the energy of RW phonon at the zone boundary (for dispersion curves of the RW phonon corresponding to the Xe(111) surface see Fig. 11 of Ref. \onlinecite{Xe111}). The same effect occurs also in connection with excitation of LR and SH modes but is correspondingly smaller due to their much weaker coupling to the scattered He atom.       

As the energy of the incident He atom is increased from $E_{i}=10.4$ meV to $E_{i}= 25$ meV, the scattering regime changes from a two-phonon to a three-phonon one, as signified by the increase of the Debye-Waller exponent from the value $2W=2.2$ to the value $2W=3.5$. 
In the three-phonon inelastic He atom scattering processes the constraints on the wavevectors and energies of the emitted and absorbed phonons are much less severe and therefore the three-phonon structure appears much smoother than the two-phonon one. This is a general trend which holds for higher order phonon processes as well. 
The described situation is illustrated in the lower LHS panel of Fig. \ref{vas10_2}. 
Here the three-phonon structure, whose manifestation are the shoulders denoted by M$_{\pm 3}$ in the main body of the multiphonon spectrum, is the dominant one regarding the total spectral weight although the single- and two-phonon features still contribute significant components to the total intensity. 

Lastly, as the incident projectile energy is increased to $E_{i}=50$ meV, the DWE reaches the value $2W=7.2$ and, accordingly, the calculated spectrum exhibits very little structure on top of a dominantly Gaussian-like shape typical of the multiphonon scattering regime. This is illustrated in the lower RHS panel in Fig. \ref{vas10_2}. The elastic peak, which is solely of the multiphonon origin, appears as a shoulder whereas the spectral maximum can be interpreted as a combination of many-phonon exchange processes resulting in the mean energy transfer of the order $\Delta\bar{E}=-3$ meV. 
Thus, the doubling of the DWE from 3.5 to 7.2 in going from the LHS to the RHS panel in Fig. \ref{vas10_2} has resulted in a practically complete removal of the one- and few-phonon structures from the scattering spectra. 
This should not come as a surprise since the Gaussian spectral shape should always be expected from a distribution describing unitarized probabilities of uncorrelated scattering events in the limit of validity of the Stirling formula for approximating the factorials. Indeed, the last value of the DWE, viz. $2W=7.2$, lies at the lower boundary of this limit.

Finally, it may be of interest to investigate the validity of the Gaussian approximation for the multiphonon scattering formulae as given by Eqs. (73)-(75) of Ref. \onlinecite{HAS} and more accurately by Eqs. (291)-(293) of Ref. \onlinecite{plrep}. 
To this end we have modelled the scattering spectrum for $E_{i}=80$ meV by the exact EBA expression from Eq. (10) of Ref. \onlinecite{Xe111} and the Gaussian limit from Eq. (291) of Ref. \onlinecite{plrep}. The results are shown in Fig. \ref{vas10_3} and they illustrate that the Gaussian approximation is an excellent one under these scattering conditions and for the set of parameters characterising the  He-Xe(111) interaction\cite{Xe111}.

In summary, we have presented a detailed theoretical assessment of the energy and parallel momentum resolved scattering spectra of thermal energy He atoms inelastically scattered from surface phonons of a simple benchmark system Xe(111). The (111) surface of the van der Waals xenon crystal supports all three surface phonons, viz. the Rayleigh wave (RW), the longitudinal resonance (LR) and the shear horizontal (SH) mode, of which the RW due to its dominantly perpendicular polarization couples most strongly to the scattered He atoms. 
Hence, analogously as in the single phonon scattering regime, this mode should also give a dominant contribution to the multiple phonon exchange processes in the few- and multiphonon scattering regime. By varying the incoming energy of the scattered He atoms in expressions for the mean number of exchanged phonons $\bar{n}$ (equal to the Debye-Waller exponent $2W$) and for the inelastic scattering intensities we have numerically simulated the HAS-TOF spectra pertinent to the He$\rightarrow$Xe(111) collision system in the various scattering regimes.    
We have shown by comparing experimental and theoretical results that in the transition from one phonon- to a two-phonon scattering regime the well resolved one-RW and one-LR phonon peaks represent the dominant structure both in the theoretical and measured spectra, but also that some extra peaks observed in experiments for incident He atom energy $E_{i}=10.5$ meV can be interpreted in terms of the "zone edge focused" two-phonon processes. 
By increasing the incident He atom energy and thereby entering the regime of higher order phonon excitation processes ($3\leq \bar{n} \leq 6$) we have shown that, contrary to the popular belief, the many-phonon contribution to the overall scattering spectrum may exhibit a relatively rich and complicated structure. Therefore, in this regime and before the multiphonon Gaussian limit is reached for $\bar{n}\geq 7$, the many-phonon contribution does not by any means represent a featureless background. Finally, the existence of a Gaussian limit is demonstrated by the calculations and comparisons of the exact EBA expression with the functional Gaussian limit of the same EBA multiphonon scattering spectrum for $E_{i}=80$ meV for which the mean number of exchanged phonons  $\bar{n}=10.9$ is large enough to fullful the conditions for a true semiclassical multiphonon limit. 

Although our findings and conclusions have been illustrated on a system that exhibits relatively  simple vibrational properties, they should be of a more general validity. Hence they may serve as a guideline in pinpointing and detecting the peculiarities of the multiphonon structures also in other systems that have been studied by HAS-TOF spectroscopy in the various scattering regimes.

The authors would like to thank A.P. Graham and J.P. Toennies for communication and discussion of the experimental data on He atom scattering from Xe(111), and G. Witte and Ch. W\"{o}ll for useful comments on the properties of surface phonon density of states. 
 This work has been supported in part by the Joint National Science Foundation grant JF 133 and the German-Croatian Collaboration Project-No.: KRO-007-97.

\protect\begin{figure}
\protect\caption{Full noisy line: experimental time-of-flight spectrum for He atom scattering from Xe(111) surface along the $\bar{\Gamma}\bar{\mbox{K}}$ direction of the first SBZ for the scattering conditions given in the inset. Full thick line: theoretical total scattering spectrum calculated in the EBA formalism (see main text) for the same set of scattering parameters. Dashed line: theoretical many-phonon spectrum (total minus one-phonon component) which in the present scattering regime is dominated by the two-phonon contributions. The symbols RW and L denote the one-phonon peaks that originate from excitation and annihilation of the  Rayleigh wave and longitudinal resonance modes. The no-loss or zero energy transfer line in the experimental spectrum bears additional weight due to elastic He atom scattering from surface defects.     } 
\protect\label{vas10_1}
\protect\end{figure}

\protect\begin{figure}
\protect\caption{Theoretical simulations of the time-of-flight spectra for He atom scattering from the Xe(111) surface at temperature $T_{s}=40$ K. The incident angle of He atoms is the same in all four panels, $\Theta_{i}=53.9^{0}$, and the incident energy $E_{i}$ is varied. Full line: total scattering spectrum calculated in the EBA formalism for incident energy as denoted in the inset of each panel. Dashed line: many-phonon contribution to the total spectrum (total minus one phonon component) whose spectral weight increases with the increase of the He  incident energy and the Debye-Waller exponent (DWE). The symbols RW and L denote the peaks that originate from excitation and annihilation of the Rayleigh wave and longitudinal resonance phonons. The symbols M denote the multiphonon structure discussed in the main text.    
 }
\protect\label{vas10_2}
\protect\end{figure}

\begin{figure}
\caption{ Calculated multiphonon limit of the spectrum of He atoms scattered from phonons of the Xe(111) surface for the parameters denoted in the inset. Full line is the result of exact calculations in the EBA formalism. Open squares represent the results of the Gaussian limit of the EBA scattering spectrum formulae for the same set of scattering parameters. The Gaussian parameters are $\Delta\bar{E}=-3.4$ meV (mean energy transfer) and $\sigma=10.6$ meV (Gaussian spectral width).}
\label{vas10_3}
\end{figure}


\begin{references}



\bibitem{Legge}H. Legge, J.P. Toennies and J. L\"{u}decke in 
{\it Rarefied Gas Dynamics 19 (1994)}, Vol.
II, edited by J. Harvey and G. Lord (Oxford Science Publications), 
p. 988; J.P. Toennies, {\it ibid}, p. 921.

\bibitem{Hulpke} See the articles in: {\em Helium Atom Scattering from 
Surfaces}, Springer Series in Surface Sciences Vol. {\bf 27}, edited by 
E. Hulpke (Springer, Berlin, 1992).

\bibitem{comment} B. Gumhalter and A. Bili\'{c}, Surf. Sci. 
 370(1997)47.

\bibitem{Hofmann} F. Hofmann, J.P. Toennies and J.R. Manson, J. Chem. Phys. 106(1997)1234.


\bibitem{sibener} K.D. Gibson and S.J. Sibener, Phys. Rev. Lett. 55(1985)1514; K.D. Gibson and S.J. Sibener, Faraday Discuss. Chem. Soc. 80(1985)203; K.D. Gibson, S.J. Sibener, B.M. Hall, D.L. Mills and 
J.E. Black, J. Chem. Phys. 83(1985)4256; K. Kern, R. David,, R.L. Palmer and G. Comsa, Phys. Rev. Lett. 56(1986)2823; K. Kern, R. David, R.L. Palmer and G. Comsa, Surf. Sci.
 175(1986)L669; P. Zeppenfeld, U. Becher, K. Kern and G. Comsa, J. Elec. Spectr. Rel. Phenom. 54(1990)265; L.W. Bruch, A. P. Graham and J.P. Toennies, 
Mol. Phys. 95(1998)579; L.W. Bruch, A.P. Graham and J.P. Toennies, J. Chem. Phys. 112(2000)3314.





\bibitem{Xemulti} A.P. Graham, M.F. Bertino, F. Hofmann, J.P. Toennies,
and Ch. W\"{o}ll, J. Chem. Phys. 106(1997)6194; J. Braun, D. Fuhrmann, A. \v{S}iber, B. Gumhalter
and Ch. W\"{o}ll, Phys. Rev. Lett. 80(1998)125; A. \v{S}iber, B. Gumhalter, J. Braun, A.P. Graham,
M.F. Bertino, J.P. Toennies, D. Fuhrmann and Ch. W\"{o}ll, Phys. Rev. B59(1999)5898.


\bibitem{WitteCO} A.M. Lahee, J.P. Toennies and Ch. W\"{o}ll, Surf. Sci. 177(1986)371;
G. Witte, H. Range, J.P. Toennies and Ch. W\"{o}ll, J. Electron Spectr. Rel. Phenom. 64/65(1993)715; 
G. Witte, J.P. Toennies and Ch. W\"{o}ll, Surf. Sci. 323(1995)228.

\bibitem{Witte} J. Ellis, A. Reichmuth, J.P. Toennies and G. Witte, J. Electron Spectr. Rel. Phenom. 64/65(1993)725.


\bibitem{SRL} B. Gumhalter, K. Burke and D.C. Langreth, Surf. Rev. Lett. 1(1994)133.

\bibitem{recovery}B. Gumhalter, A. \v{S}iber and J.P. Toennies, 
Phys. Rev. Lett. 83(1999)1375.

\bibitem{plrep} B. Gumhalter, Physics Reports 351/1-2(2001)1.


\bibitem{DWF} B. Gumhalter, Surf. Sci. 347(1996)237.


\bibitem{Xe111} A. \v{S}iber, B. Gumhalter, A.P. Graham and J.P. Toennies, Phys. Rev. B63(2001)115411.




\bibitem{HAS} A. Bili\'{c} and B. Gumhalter, Phys. Rev. B52(1995)12307.







\end{references}
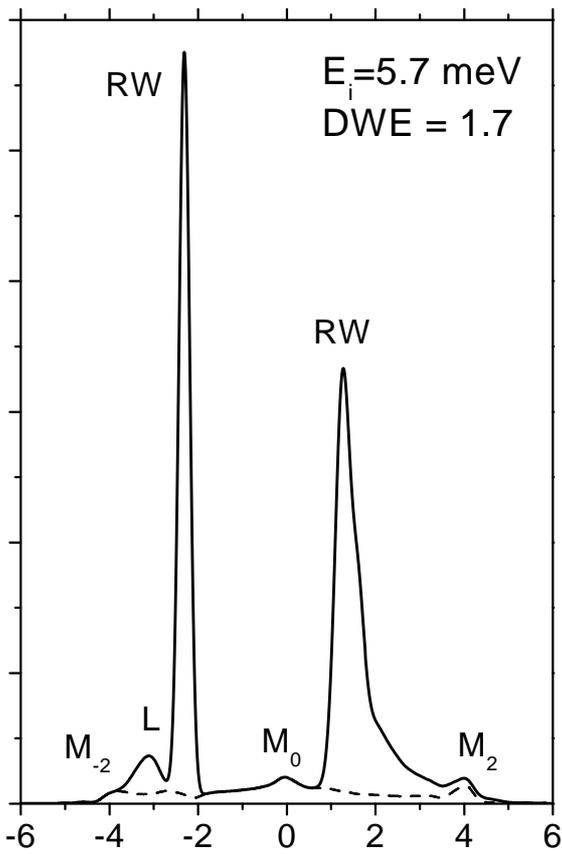
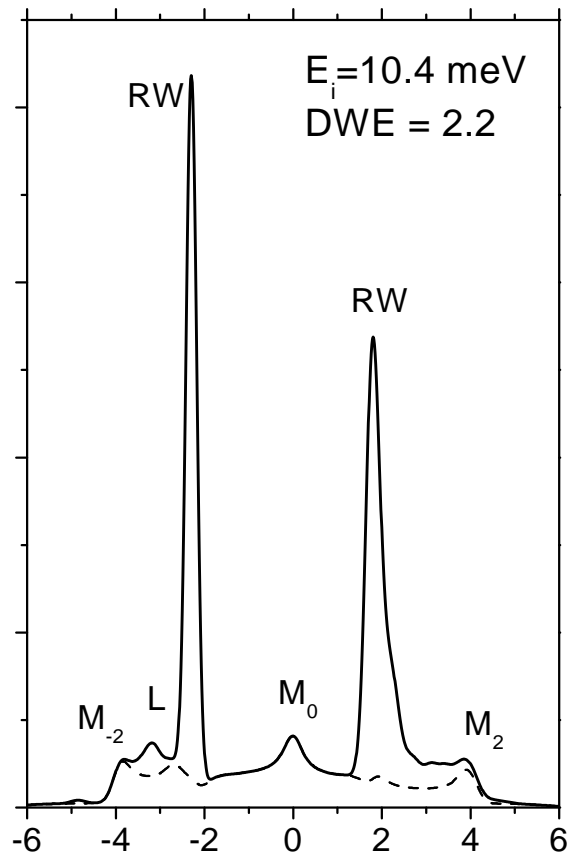
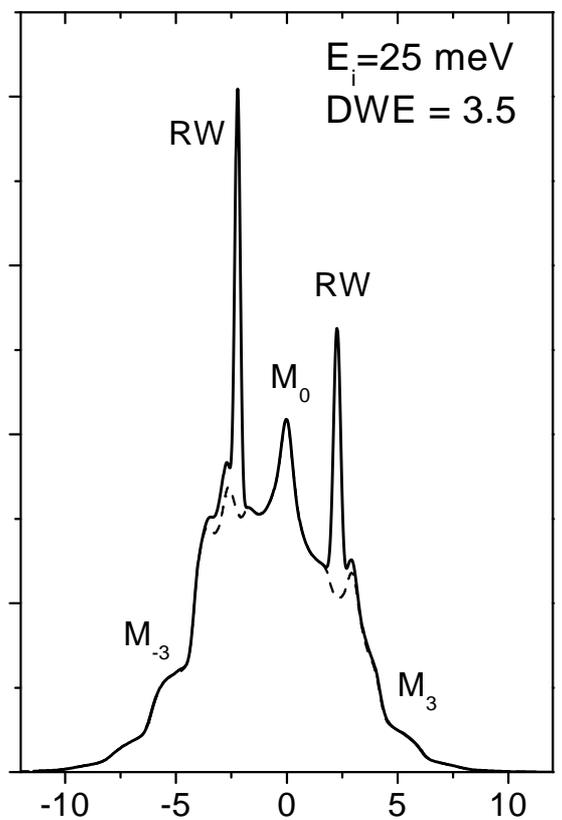
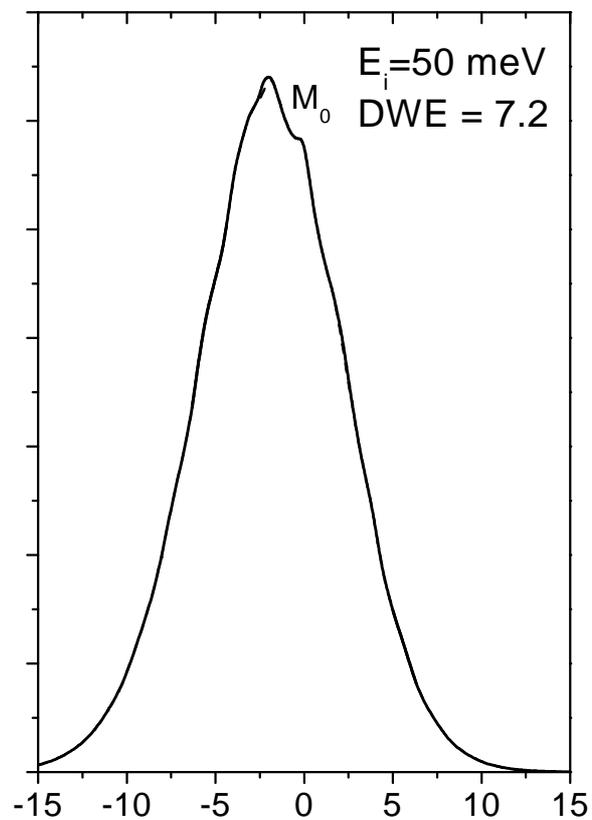
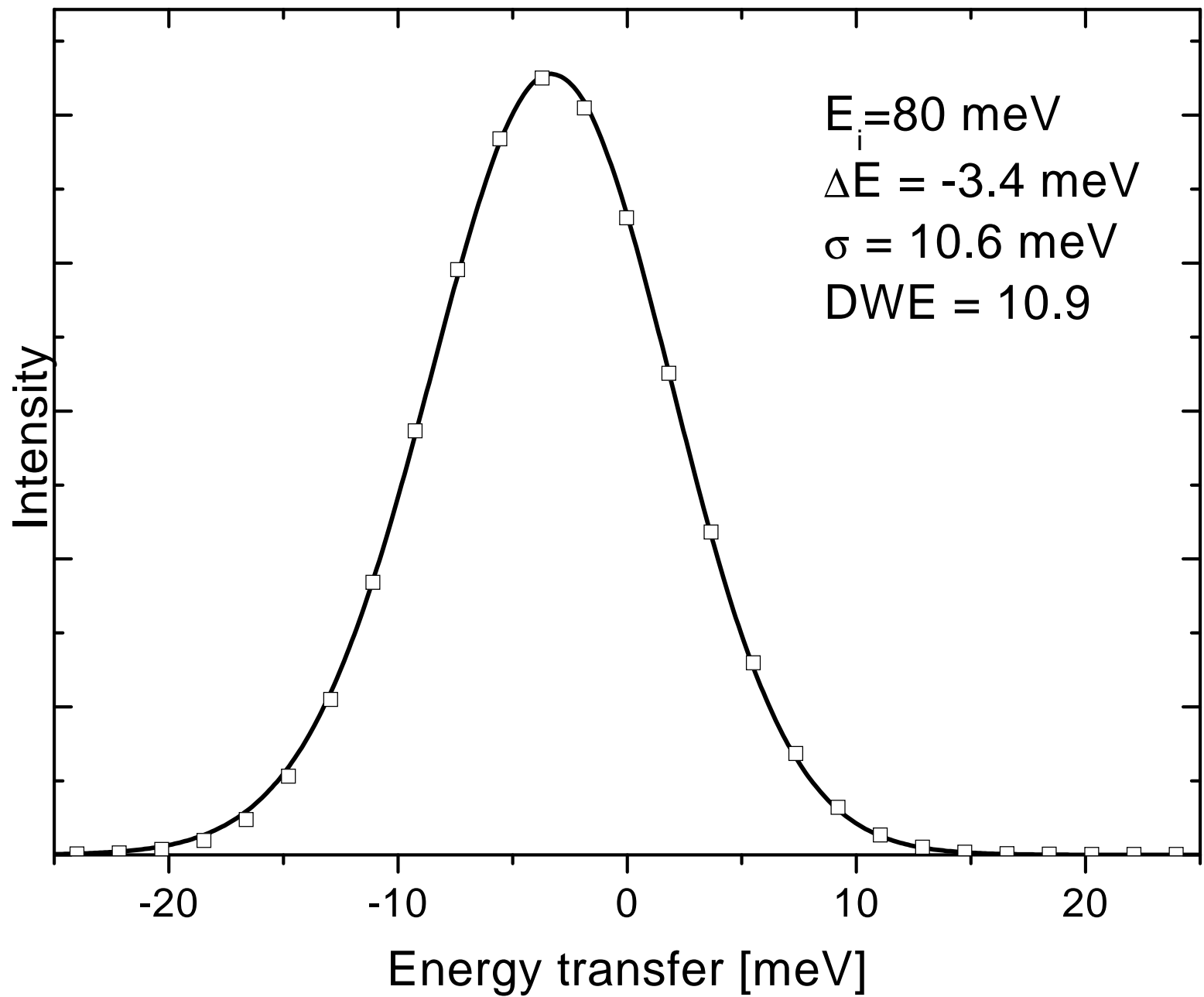
\end{document}